\documentclass[conference]{IEEEtran}
\ifCLASSINFOpdf
  \usepackage[pdftex]{graphicx}
\else
  \usepackage[dvips]{graphicx}
\fi
\usepackage{url}

\begin{document}
%
\title{Growing Self-organized Design of \\
Efficient and Robust Complex Networks}

\author{\IEEEauthorblockN{Yukio Hayashi}
\IEEEauthorblockA{School of Knowledge Science\\
Japan Advanced Institute of Science and Technology\\
Nomi, Ishikawa 923-1292\\
Email: yhayashi@jaist.ac.jp}
}


%


\maketitle

\begin{abstract}
A self-organization of 
efficient and robust networks is important for a future design 
of communication or transportation systems, 
however both characteristics are incompatible in many real 
networks.
Recently, 
it has been found that the robustness of 
onion-like structure with positive degree-degree correlations
is optimal against intentional attacks.
We show that, 
by biologically inspired copying, 
an onion-like network 
emerges in the incremental growth 
with functions of proxy access and 
reinforced connectivity on a space. 
The proposed network consists of 
the backbone of tree-like structure by copyings 
and the periphery by adding shortcut links between low degree nodes
to enhance the connectivity.
It has the fine properties 
of the statistically self-averaging 
unlike the conventional duplication-divergence model, 
exponential-like degree distribution without overloaded hubs, 
strong robustness against both malicious attacks and random failures, 
and the efficiency with short paths counted by the number of 
hops as mediators and by the Euclidean distances. 
The adaptivity to heal over and to recover the performance of 
networking is also discussed for a change of environment in such 
disasters or battlefields on a geographical map.
These properties will be useful for a resilient and scalable 
infrastructure of network systems even in 
emergent situations or poor environments.
\end{abstract}


%
\IEEEpeerreviewmaketitle

\section{Introduction} \label{sec1}
Self-organizations with some outstanding properties of 
no central control, emerging structures, resulting complexity, 
and high scalability \cite{Dressler07} 
appear in natural, social, and technological systems.
To make a structure autonomously, 
movements or interactions of objects (materials or information)
are necessary, therefore naturally induce their flows and 
form a network as the base. 
There are several fundamental mechanisms: 
preferential attachment \cite{Barabasi99}, 
copying \cite{Sole02, Satorras03}, 
survival \cite{Tero10, Hayashi12a}, 
subdivision 
(fragmentation) \cite{Doye05, Zhou05, Hayashi09a, Hayashi09b}, 
or aggregation \cite{Alava05}, 
for generating networks in the interdisciplinary research fields of 
physics, biology, sociology, and computer science.
They are summarized in Table \ref{table_self-org}.
However, even the state-of-the-art science and technology 
is still far from fully understanding of the potential for 
applying to a design of efficient and robust networks, 
while our daily life strongly depends on network infrastructures in 
energy supply, communication, transportation, economic, 
ecological, and biological (nervous or genetic) systems.

As one of the most elucidated evidences, 
it has been found \cite{Dorogovtsev04, NBW06, Newman10, Choen10}
that many real networks in social, biological, 
and information systems have a {\it scale-free} (SF)
structure whose degree distribution follows a power-law; 
e.g. the networks of coauthors, movie actors, protein interactions, 
food webs, Internet, and WWW are included in this class \cite{Albert02}.
The SF network consists of many low degree nodes and a few 
very large degree nodes. 
The key mechanism for network generation 
is the preferential attachment
\cite{Barabasi99} called as {\it rich-get-richer} rule. 
In other words, 
people ($=$ nodes) tend to do economic trade with richer persons 
($=$ linking to large degree nodes)
to get more money in selfishness.  
The SF networks have the efficient {\it small-world} (SW) 
property \cite{Watts98} 
that the path length between any two nodes is short 
even for a large network size, 
however they have an extreme 
vulnerability against intentional attacks \cite{Albert00}. 
Thus, 
the efficiency and the robustness are incompatible in 
many real networks.
In addition, the percolation analysis in statistical physics 
has recently found \cite{Buldyrev10, Huang11, Huang12}
that the vulnerability increases in interdependent networks: 
realistic modern systems consisting of mutually related vulnerable 
networks in power-grid, communication, 
transportation, economic-trading, and so on.
Disasters or terrorism are no longer unusual, 
moreover the threat to induce a malfunction 
becomes more serious.
Because the above current networks are extremely vulnerable
in the interdependency, 
nevertheless to maintain the connectivity is the most fundamental 
function as network.

Such a fat-tail distribution as power-law 
is also emerged mathematically 
by other quite random and unselfish mechanisms,
e.g, multiplicative process: a quantity $X_{t}$ at time $t$ is given 
by the product 
$X_{t} = \left( \Pi_{s=1}^{t-1} r_{s} \right) X_{0}$ 
\cite{Mitzenmacher04}, 
in which a temporal accumulation of random growth rate 
$0 < r_{s} < 1$ contributes 
to make the inequality even for the 
uniform randomness at each time step. 
Because the distribution follows a log-normal 
from the central limit theory for
$\log X_{t} = \sum_{s} \log r_{s} + \log X_{0}$. 
Similar distributions have been obtained in 
a simple model of random copying among individuals in the evolution 
for studying cultural change \cite{Neiman95,Bentley03,Bentley04}. 
The distributions consist of many uncommon
variants and a very few common variants, 
where cultural variants means first names, journal
citations, decorative motifs on archaeological pottery, and patent
citations in U.S. 
Thus, random copying may induce a key 
mechanism to self-organize a complex structure 
depending on a probabilistic selection history, 
however it is not a network model. 
On the other hand, the biologically inspired 
duplication process has been so far considered to be fundamental
in a model of protein-protein interaction networks
\cite{Sole02, Satorras03}.
 It suggests that 
{\bf complex network structures can be generated by a simple
random process}.

In this paper, 
we consider a self-organized design of 
efficient and robust networks by biologically inspired 
copying.
In particular, we focus on the robust network structure
 with positive degree-degree correlations 
\cite{Schneider11, Herrmann11, Tanizawa12} 
and an incrementally growing mechanism.
The self-propagation in maintaining the robust structure 
is particularly important for the scalability 
of network system without degrading the performance.

The organization of this paper is as follows. 
In Sec. \ref{sec2}, we briefly review related network models to ours.
In Sec. \ref{sec3}, 
we propose a basic model 
for understanding the fundamental mechanism of network 
self-organization
from a viewpoint in complex network science 
\cite{Dorogovtsev04, NBW06, Newman10, Choen10} which aims to 
elucidate the universal properties and the generation rule of 
networks.
We show the important properties of our model 
for the robustness of connectivity 
and the emergent functions.
In Sec. \ref{sec4}, we consider an incrementally growing mechanism, 
and investigate the strong robustness 
and the efficiency for path lengths. 
We emphasis that the performance of our proposed networks 
for both the robustness and the efficiency 
does not degrade rather rises through the growth. 
In Sec. \ref{sec5}, we discuss the possibilities and the issues 
for applications especially in communication or transportation systems.
The adaptivity for changing environment in disasters or battlefields 
is also discussed. 
In Sec. \ref{sec6}, we summarize there results.

\begin{table}
\centering
\scalebox{1.3}{
\begin{tabular}{l|l|l}
Model    & Mechanism & Reference \\ \hline
network  & preferential & \cite{Barabasi99} \\
         & attachment & \\ \hline
cultural change, & copying or & \cite{Neiman95,Bentley03,Bentley04} \\
network  & duplication & \cite{Sole02, Satorras03} \\ \hline
trail,   & survival & \cite{Helbing97} \\
network  &          & \cite{Tero10, Hayashi12a} \\ \hline
cracking, & fragmentation or & \cite{Nagel08, Krapivsky10} \\ 
network & subdivision & \cite{Doye05, Zhou05, Hayashi09a, Hayashi09b} \\ \hline
cluster of mass, & aggregation & \cite{Krapivsky10} \\ 
network & & \cite{Alava05} \\ \hline
\end{tabular}
}

\vspace{2mm}
\caption{Typical self-organization mechanisms.}
\label{table_self-org}
\end{table}

\section{Related Works in Complex Network Science} \label{sec2}
It is a reasonable assumption that 
a biological network grows according to a simple mechanism 
based on random copying 
which differs from the preferential attachment \cite{Barabasi99} 
by selecting large degree nodes in selfishness. 
Thus, in the duplication-divergence (D-D) model
\cite{Sole02, Satorras03}, 
a new node is added at each time step, 
and duplicately creates links 
to connected neighbor nodes of a randomly chosen node.
Some links in the duplication are deleted with a probability $\delta$.
Without using degrees for the selection of linked nodes, 
a distribution of power-law with exponential cutoff emerges 
in the D-D model.
However, there is a serious problem that 
the models by pure duplication without deletion of $\delta=0$
and by D-D with weak deletion of small $\delta < 1/2$ 
have a singular property called as 
{\it non-self-averaging} \cite{Satorras03, Ispolatov05};
In the pure duplication \cite{Ispolatov05}, 
bipartite graphs $K_{j,N-j}$ in any combinations of positive
integers $j$ and $N-j$ are generated with equiprobability, 
the degrees of $j$ and $N-j$ distribute uniformly.
The statistical quantities such as the degree distribution and 
the total number $j (N-j)$ 
of links in a sample of the networks 
have {\bf no meaning to characterize the topological structure},
since the range between the minimum $N-1$ at $j=1$
and the maximum $(N/2)^{2}$ at $j=N/2$
diverges for a large size $N$. 
In addition, 
the attack vulnerability remains in the D-D model because of the 
SF-like network. 

\begin{figure}[htb]
\begin{center}
  \includegraphics[height=55mm,angle=-90]{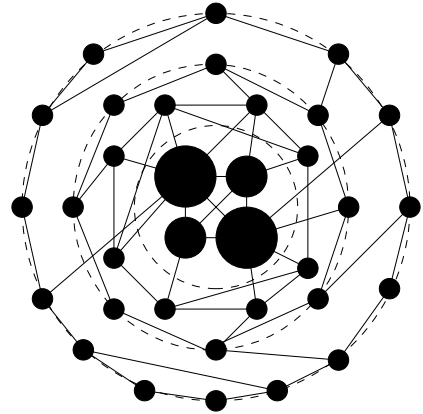}
\caption{Illustration of onion-like network structure, in which 
nodes with a same degree are set on each concentric circle 
denoted by a dashed line. The solid line denotes a link 
between two nodes.
The size of node is proportional to its degree.}
\label{fig_onion_struct}
\end{center}
\end{figure}

On the other hand, 
{\bf the optimal structure against the targeted attacks to hub nodes}
has been shown numerically \cite{Schneider11, Herrmann11} and 
analytically \cite{Tanizawa12} in a SF network.
It is referred to {\it onion-like structure} (see Fig. \ref{fig_onion_struct}) 
from the character of connectivity 
with positive degree-degree correlations, 
which correspond to a natural tendency of homophily: 
nodes are more likely connect to other nodes that are similar to them.
Under a given degree distribution, 
an onion-like network with the nearly optimal robustness 
can be generated by rewirings 
to enhance the degree-degree correlations \cite{Wu11}. 
Thus, the onion-like structure is applicable 
to a network with any degree distribution. 
For example, after the rewirings, 
the attack vulnerability is decreased 
for a power-law degree distribution in a SF network generated by the 
preferential attachment or other methods.
However, 
{\bf the entire rewiring process discards already existing links, 
it is not effective utilization when a network is growing 
in a realistic situation}.
Moreover, in spite of the improvement of robustness, 
the positive degree-degree correlations in a SF network 
tend to induce 
longer path lengths \cite{Tanizawa13}, which are undesirable 
for the efficient communication or transportation.
We should not persist SF networks, 
and study other candidates for future design of networks, 
especially in considering self-organization mechanisms.

\section{Basic Properties} \label{sec3}
In this section, 
we consider a self-organization of onion-like network, 
and show the tolerant and topological properties of robustness, 
degree distribution, and degree-degree correlation. 
To study the connectivity in a network structure 
independently of applications is important for clarifying 
 the basic property.

\subsection{Network model}
\begin{figure}[htb]
\begin{center}
  \includegraphics[height=45mm]{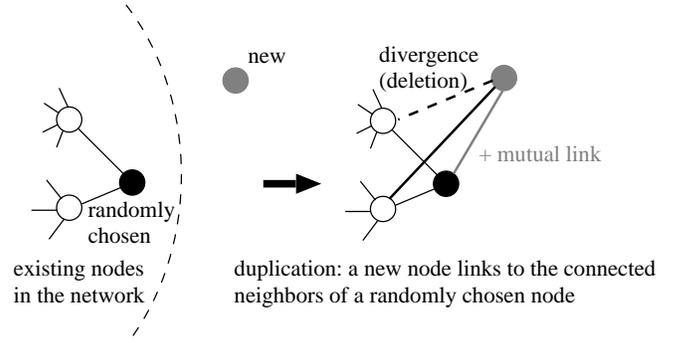}
\caption{Basic process in constructing a network.}
\label{fig_model}
\end{center}
\end{figure}

We propose a modification of the D-D model 
\cite{Sole02, Satorras03, Ispolatov05} without mutation.
Mutations by connecting randomly chosen two nodes are 
unnecessary for generating an onion-like network.
The basic processes of network construction 
by using only local information are as follows 
(see Fig. \ref{fig_model}). 

\begin{description}
  \item[Step 0:] Set an initial configuration 
    (e.g. connected two nodes).
  \item[Step 1:] At each time step $t = 1, 2, \ldots$, 
    a new node is added.
    The new node $i$ connects to a randomly chosen node 
    and to the neighbor nodes $j$ 
    with a probability $(1 - \delta) \times p$ \cite{Wu11}, 
    \begin{equation}
      p = \frac{1}{1 + a \mid k_{i} - k_{j} \mid},
      \label{eq_ass_link}
    \end{equation}
    where $\delta$ is a rate of link deletion, 
    $a \geq 0$ is a parameter, 
    and $k_{i}$ and $k_{j}$ denote the degrees of nodes $i$ and $j$.
  \item[Step 2:] The above processes are repeated 
    up to a given size $N$ 
    in prohibiting self-loops at a node 
    and duplicate connections between two node.
\end{description}

In Step 1, $k_{i}$ is temporary 
set as $(1 - \delta) \times$ the degree of the chosen node, 
since the degree of new node is unknown in advance because of 
the stochastic process. 
When degrees $k_{i}$ and $k_{j}$ 
are close, the two nodes $i$ and $j$ are connected with high 
probability. The case of $a=0$ without the correlation effect 
corresponds to the conventional D-D model
except the mutual link between new node and chosen node.
We set $a = 0.3$.

\begin{figure}[htb]
\begin{center}
  \includegraphics[height=91mm,angle=-90]{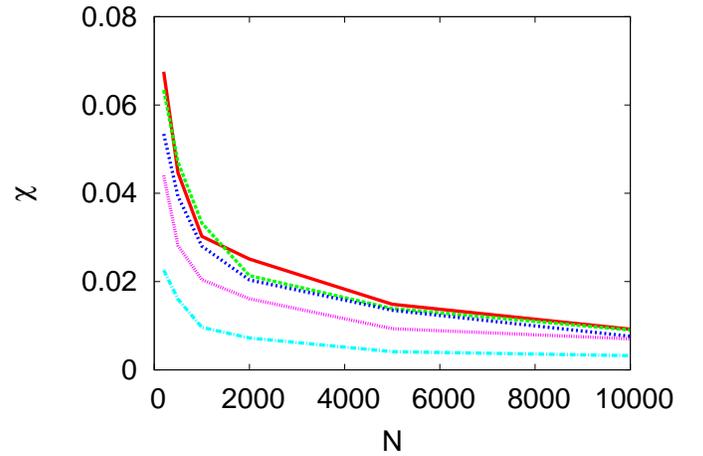}
\caption{Self-averaging with a convergence over $100$ realizations.
The red, green, blue, magenta, and cyan lines correspond to the cases 
of $\delta = 0.1$, $0.3$, $0.5$, $0.7$ and $0.9$, respectively.}
\label{fig_self-ave}
\end{center}
\end{figure}

\begin{figure}[htb]
\begin{center}
  \includegraphics[height=126mm]{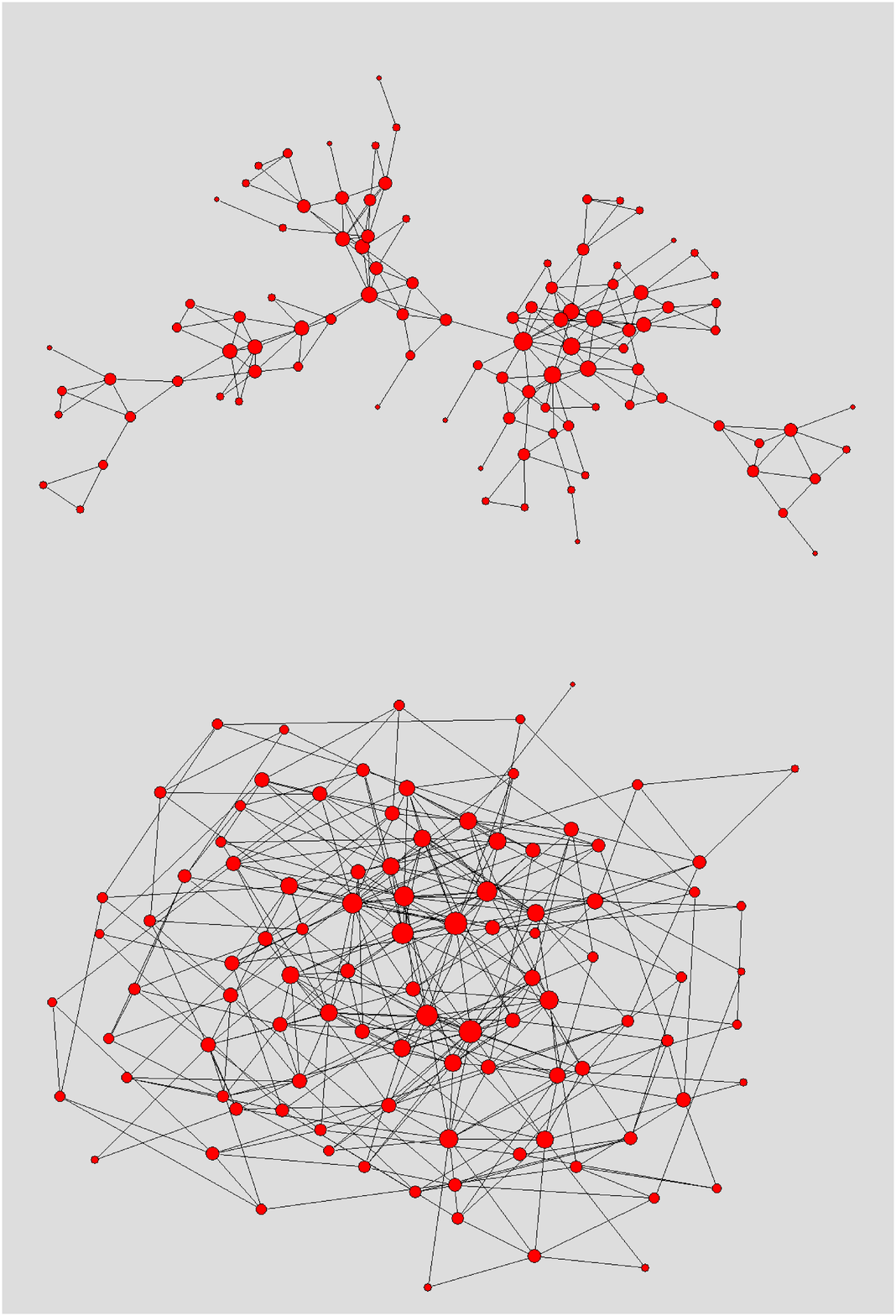}
\caption{Visualization examples for $N=100$. 
(Top) a tree-like network by copyings for $\delta = 0.5$.
(Bottom) the extended onion-like network by adding shortcut links.
The sizes of nodes are proportional to their degrees.}
\label{fig_vis_ex}
\end{center}
\end{figure}

Figure \ref{fig_self-ave} shows that our proposed networks satisfy 
a good property of the self-averaging unlike the conventional 
D-D model: 
the statistical deviation 
$\chi \stackrel{\rm def}{=} \sqrt{\langle M^{2} \rangle
- \langle M \rangle^{2}} / \langle M \rangle$ 
for the total number $M$
of links converges to zero for a large size $N$. 
Here, $\langle \;\; \rangle$ denotes the statistical mean 
(expectation). 
We remark a reason of the self-averaging by that 
a sequence of complete graphs is generated in our case of pure 
duplication of $\delta = 0$ instead of bipartite graphs in the 
conventional D-D model \cite{Ispolatov05}.
Therefore, with a small change from the conventional D-D model 
by adding the mutual links, 
the tendency of dense connections 
(whose extreme case is a complete graph)
induces a core of connected high degree nodes. 
In addition, since there is an effect of preferential attachment 
to the random neighbors \cite{Yang13, Colman13}, large degree nodes 
$i$ and $j$ tend to be connected together
when the chosen node has a large degree.
The positive degree-degree correlations is suitable to improve the 
robustness, especially for the malicious attacks
\cite{Schneider11, Herrmann11}.
However, such correlations between low degree nodes are weak in 
the tree-like structure 
as shown in the top of Fig. \ref{fig_vis_ex}.
Since a low degree node dangles from a higher degree node, 
the dangling part is easy to be disconnected by node removals. 
We remark that the majority is low degree nodes.
Thus, we consider the addition of shortcut links 
\cite{Watts98} in Step 3.

\begin{description}
  \item[Step 3 :] After Step 2, some shortcut links 
    between randomly chosen nodes $i$ and $j$ are added 
    with the probability of Eq.(\ref{eq_ass_link}). 
    The number of shortcut links are given by $p_{sc} M$, 
    where $p_{sc}$ is an adding rate of shortcut and $M$ is the 
    total number of links in the original tree-like network
    generated for the size $N$.
\end{description}
The bottom of Fig. \ref{fig_vis_ex} shows an onion-like structure
by adding shortcut links after Step 2.
In the next subsection, as a basic property, 
we will separately discuss the effects of copyings and adding shortcuts 
on the robustness in order to make them clearly.

\begin{figure}[htb]
\begin{center}
  \includegraphics[height=95mm,angle=-90]{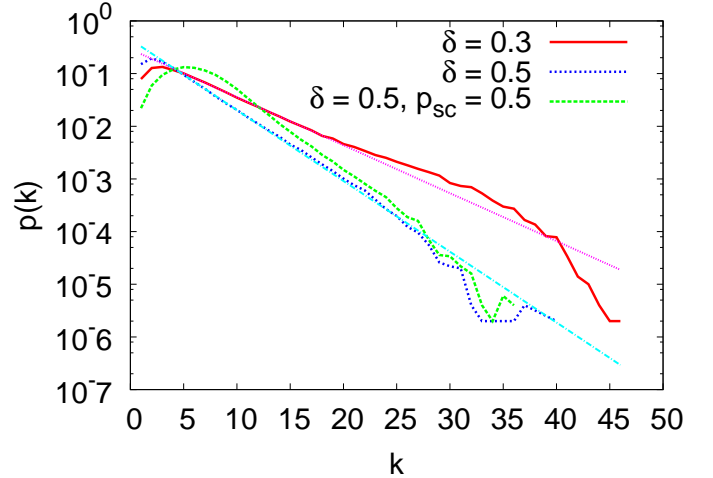}
\caption{Degree distributions over $100$ 
realizations of our proposed network for $N=5000$.
The dotted magenta and chain cyan straight 
lines are the approximations 
by exponential distributions: $0.289 \exp(-0.209 k)$ and 
$0.446 \exp(-0.309 k)$, respectively.
By adding shortcut links, the fraction of 
small degree nodes is slightly increased 
from the dotted blue line to the dashed green line.}
\label{fig_pk_approx}
\end{center}
\end{figure}

We emphasize the two important functions.
First, {\bf a new node act as a local proxy of the chosen node}
and becomes another access point for the neighbor nodes 
in an analogy of distributed computer communication systems. 
Second, {\bf the complementary added shortcuts improve the robustness}. 
It has already been shown that 
adding some shortcut links between uniform randomly chosen nodes 
improves the robustness 
in the theory for the small-world model \cite{Newman00}
and also in the numerical simulations for many networks 
\cite{Hayashi12a, Hayashi09a, Hayashi09b, Hayashi07}.
The robustness is further improved 
due to the positive degree-degree correlations in the onion-like
network as shown later, 
however the case of $\delta=0$ as the uniform random selections 
does not give the nearly optimal robustness.

In addition, 
the proposed network has an efficiency without 
maintaining huge connections. 
Figure \ref{fig_pk_approx} shows that 
the degree distributions are approximated by exponential distributions. 
Therefore, huge degree nodes do not appear in the network, 
it is suitable for avoiding both the attack vulnerability \cite{Albert00}
and the concentration of linking cost or traffic load
\cite{Goh01}. 
Note that the proposed network does not belong to a SF network, 
since the degree distribution is not a power-law but an exponential.

\subsection{Measures of robustness and degree-degree correlations}
We consider the malicious attacks, 
in which nodes are removed in decreasing order of the current 
degrees through the recalculations \cite{Holme02}.
This type of attacks is a more serious problem setting from an 
assumption of circumspect terrorism than the attacks 
without recalculations. 
Attackers tend to aim at large degree nodes, 
since the removals give considerable damage to maintaining paths 
by many disconnections emanated from the removed nodes.
We compare the robustness of the proposed network with 
the nearly optimal case 
under the same degree distribution.
The nearly optimal case is given by the rewired version 
defined as follows.
According to the stub-connection process 
until all the stubs have been used up \cite{Wu11}, 
with a probability
\[
   \frac{1}{1 + a \mid s_{i} - s_{j} \mid},
\]
the connections of randomly selected nodes $i$ and $j$ 
are repeated. 
Here, $s_{i}$ and $s_{j}$ denote the rank 
for the degrees of nodes $i$ and $j$.
A remaining small part of unpaired stubs 
is remedied by the reshuffle procedure.
In the stub-connection process, 
the initial configuration consists of all unpaired stubs: any node 
$i$ has $k_{i}$ free links (whose linked nodes are undetermined)
assigned from a given degree distribution $P(k)$.

We have two measures for investigating the robustness and 
the degree-degree correlations.
The robustness is measured by the following 
index \cite{Schneider11, Herrmann11, Wu11} 
\begin{equation}
  R = \frac{1}{N} \sum_{q = 1/N}^{1} S(q),
\label{eq_def_index-R}
\end{equation}
where 
$S(q)$ denotes the number of nodes in the giant component 
(GC: largest connected cluster) after removing $q N$ nodes by the 
malicious attacks.
$S(q)$ is monotonically decreased for increasing $q$ 
from an initial configuration of $S(0) = N$. 
The range of $R$ is $[0, 0.5]$, where $R=0$ corresponds to 
a completely disconnected network consisting of isolated nodes, 
and $R=0.5$ corresponds to the most robust network. 

The degree-degree correlation
is measured by the assortativity \cite{Newman10}
\begin{equation}
  r = \frac{S_{1} S_{e} - S_{2}^{2}}{S_{1} S_{3} - S_{2}^{2}}, 
  \label{eq_def_assort}
\end{equation}
where $S_{1} = \sum_{i} k_{i}$, $S_{2} = \sum_{i} k_{i}^{2}$
$S_{3} = \sum_{i} k_{i}^{3}$, $S_{e} = \sum_{ij} A_{ij} k_{i} k_{j}$, 
$A_{ij}$ denotes the $i$-$j$ element of the adjacency matrix. 
The right-hand side of Eq. (\ref{eq_def_assort}) is a suitable scheme 
for the numerical calculation of $r$
than the original definition \cite{Newman03a}.
The range of $r$ is $[-1, 1]$ as the Pearson correlation 
coefficient of the degree. 
Nodes with similar degrees tend to be connected as 
$r > 0$ is larger, while nodes with different degrees 
tend to be connected as $r < 0$ is smaller 
(but $|r|$ is larger).

\begin{figure}[htb]
\begin{center}
  \includegraphics[height=73mm]{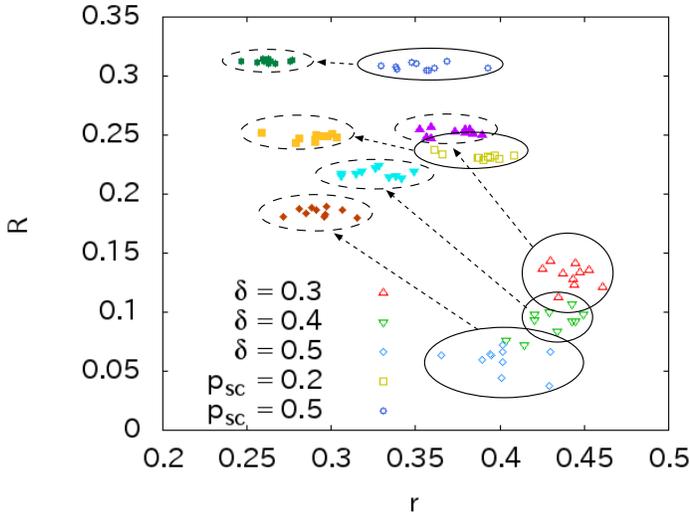}
\caption{Scatter plot of robustness index $R$ vs 
assortativity $r$ in sampling the networks 
for $N=5000$.}
\label{fig_scatter}
\end{center}
\end{figure}

Figure \ref{fig_scatter} shows a scatter plot of 
robustness index $R$ versus assortativity $r$. 
The open marks correspond to the proposed networks, 
and the filled marks correspond to 
the rewired versions at the optimal \cite{Wu11}.
We set $\delta = 0.5$ for the added versions with 
shortcut links for rates $p_{sc} = 0.2$ and $0.5$. 
The values of $R$ become larger as $\delta$ is smaller in 
the tree-like networks with more links
(denoted by open triangles and diamond marks).
While the robustness is improved by adding shortcut links 
(denoted by open square and circle marks) 
in comparison with these corresponding results in the similar 
connection density level of the average degree $\langle k \rangle$:
\begin{center}
\begin{tabular}{cccc}
$\delta = 0.5$, $p_{sc} = 0.5$ &
$\leftrightarrow$ & $\delta = 0.3$: & $\langle k \rangle \approx 6.3$ \\
$\delta = 0.5$, $p_{sc} = 0.2$ & 
$\leftrightarrow$ & $\delta = 0.4$: & $\langle k \rangle \approx 5.2$ \\
 & & $\delta = 0.5$: & $\langle k \rangle \approx 4.3$
\end{tabular}
\end{center}
and saturated around the nearly optimal for the rewired versions
(denoted by filled marks). 
The dashed arrows show the improvement for the robustness, 
and the almost flat ones mean that the proposed networks with shortcuts 
have the nearly optimal robustness.
Here, from ovals to dashed ovals, 
the decrease of $r$ is not strange for the rewired version 
to be onion-like structure with higher robustness.
Because 
it has been pointed out that onion structure and assortativity 
are distinct property  \cite{Schneider11}: 
{\it Not all assortative networks have onion structure
but all onion networks are assortative} \cite{Wu11}.

\section{Growing self-organization} \label{sec4}
We further consider an incrementally 
growing onion-like networks by simultaneous 
processes of copyings and adding shortcuts.
Because the tree-like network leaves its robustness weak in the growth 
except at the final stage by adding many shortcut links.
Thus, we modify the shortcut process as follows in Step 2.
Instead, Step 3 is omitted.

\begin{description}
  \item[Step 2':] In Step 2, at every interval $IT$, 
    shortcut links are added between randomly chosen nodes 
    $i$ and $j$ according to the probability of Eq.(\ref{eq_ass_link}).
    This process is repeated up to $p_{sc} M(t)$ links. 
    Here, $M(t)$ denotes the total
    number of links in the network at that time 
    $t = IT, 2 IT, 3 IT, \ldots$.
\end{description}
In this section, we numerically show 
the strong robustness against both attacks and failures 
and the efficiency for path lengths 
in the incrementally growing onion-like networks.
Not only the emergent structure in distributed manner but also 
the scalability without performance degradation 
are important for the self-organization, and 
don't appear in the rewiring model \cite{Wu11}.

\subsection{Robustness in the incrementally growing networks}
In order to be $\langle k \rangle \approx 6.3$ at $N=5000$ with $IT=50$, 
we chose a combination of parameters: 
$\delta = 0.4$ \& $p_{sc}=0.003$, 
$\delta = 0.5$ \& $p_{sc}=0.006$, 
$\delta = 0.6$ \& $p_{sc}=0.010$, 
$\delta = 0.9$ \& $p_{sc}=0.018$, 
and the corresponding $\delta = 0.3$ in the tree-like networks.
Since the robustness of connectivity depends on a degree distribution 
but increases 
as the average degree $\langle k \rangle$ is larger with more links
in general, 
a same level of connection density must be set to compare the 
robustness in several networks.
Note that the size is $N = t + N_{0}$ at time $t$, where $N_{0}$
denotes the initial size ($N_{0} = 2$ when the initial configuration 
is a connected two nodes).

\begin{figure}
  \begin{minipage}[htb]{.47\textwidth} 
   \centering
   \includegraphics[height=93mm,angle=-90]{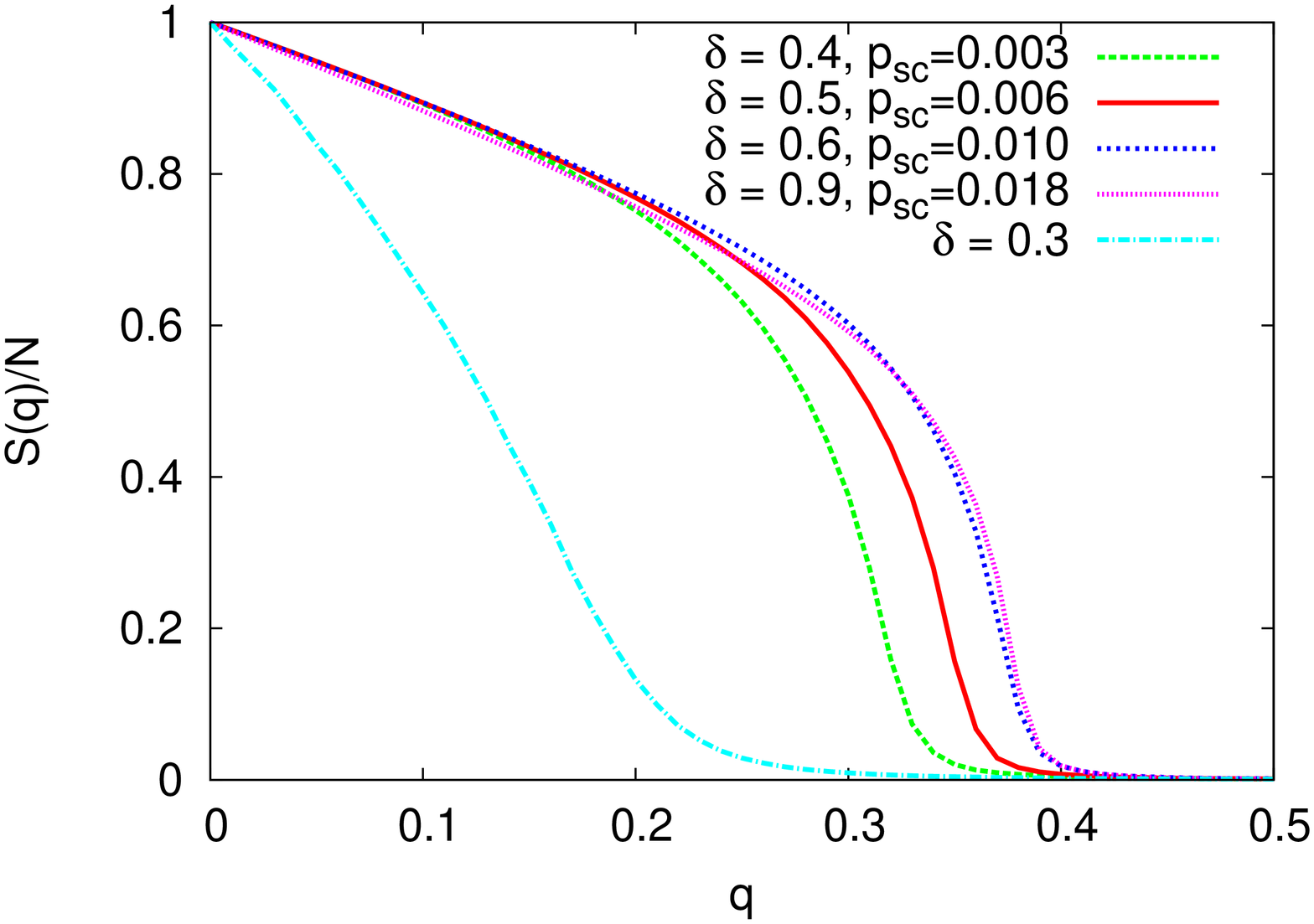}
   \begin{center} (a) \end{center}
  \end{minipage}\\
  \begin{minipage}[htb]{.47\textwidth} 
    \centering
    \includegraphics[height=93mm,angle=-90]{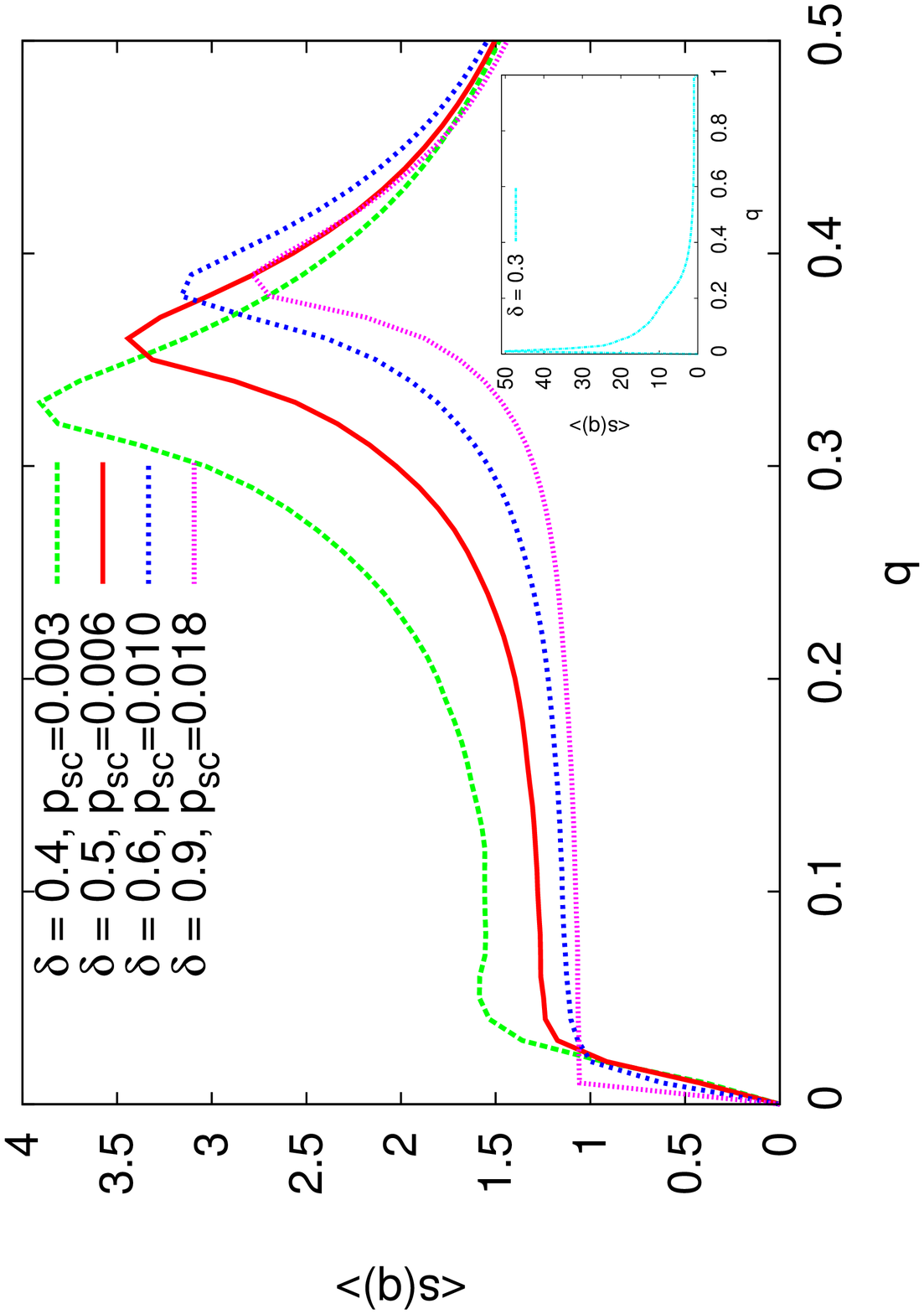}
    \begin{center} (b) \end{center}
  \end{minipage}
\caption{Attack tolerance. 
(a) Relative size $S(q)/N$ against malicious attacks for $N=5000$. 
(b) The corresponding average size $\langle s(q) \rangle$ of isolated 
clusters other than the GC. 
These results are averaged over 100 realizations $\times$ 
10 attacks to absorb 
the effect of random selection for tie-breaking of same degree.}
\label{fig_SN_attack}
\end{figure}

\begin{figure}
  \begin{minipage}[htb]{.47\textwidth} 
   \centering
   \includegraphics[height=93mm,angle=-90]{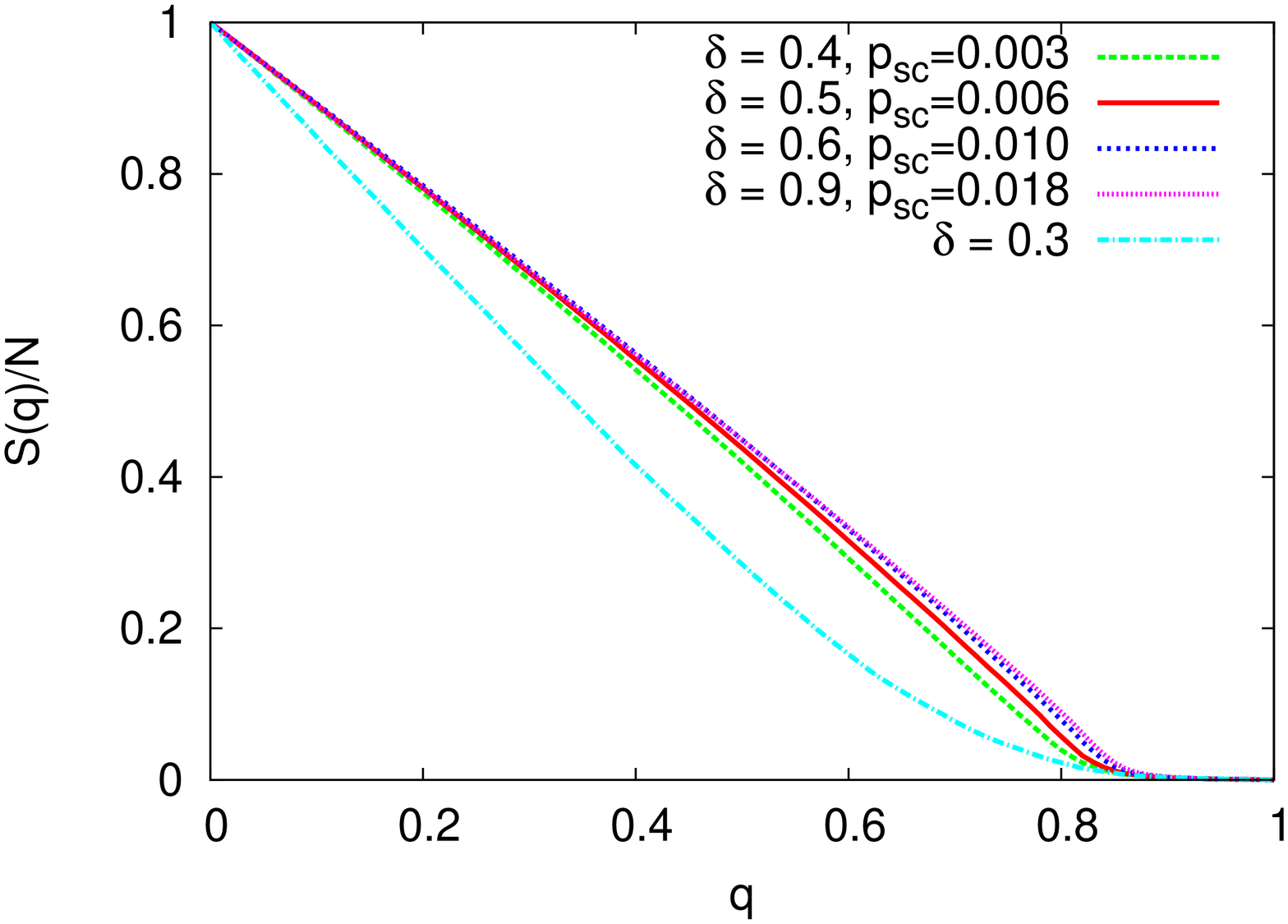}
   \begin{center} (a) \end{center}
  \end{minipage}\\
  \begin{minipage}[htb]{.47\textwidth} 
    \centering
    \includegraphics[height=93mm,angle=-90]{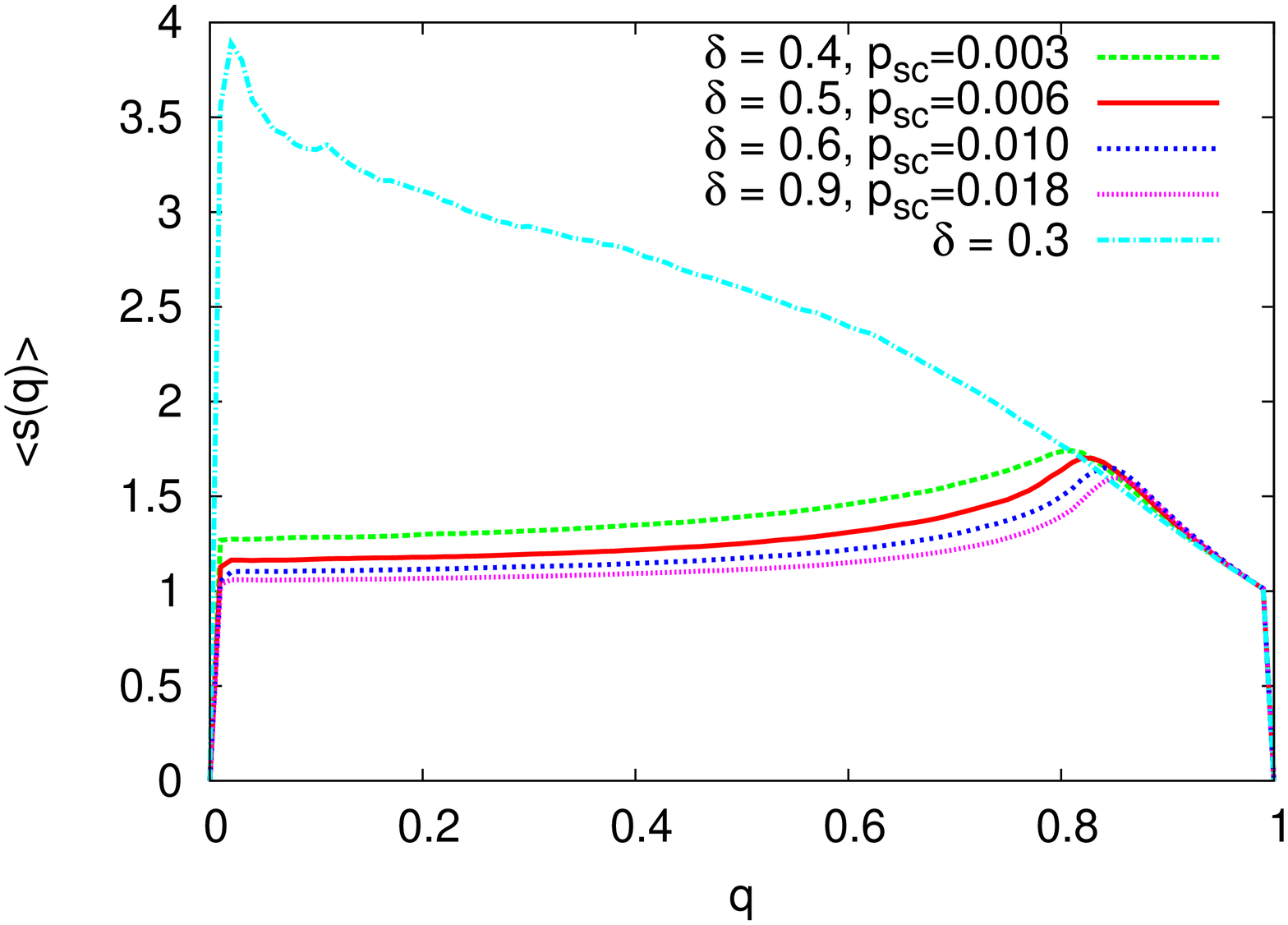}
    \begin{center} (b) \end{center}
  \end{minipage}
\caption{Failure tolerance. 
(a) Relative size $S(q)/N$ against random failures for $N=5000$.
(b) The corresponding average size $\langle s(q) \rangle$ of isolated 
clusters other than the GC. 
These results are averaged over 100 realizations.}
\label{fig_SN_failure}
\end{figure}

Figure \ref{fig_SN_attack}(a) shows 
the relative size $S(q)/N$ of nodes 
belonging to the GC versus the fraction $q$ of removed nodes 
by the malicious attacks. 
From the tree-like (denoted by cyan dashed line) 
to the onion-like (denoted by other lines) networks, 
the robustness is significantly improved, as $\delta$ is larger. 
However the difference in $\delta = 0.4 \sim 0.9$ is small.
For the breaking of the GC, 
the critical fraction $q_{c}$ is $0.3 < q_{s} < 0.4$ 
for the onion-like networks 
but very small $\ll 0.1$ for the tree-like networks 
as shown in Fig. \ref{fig_SN_attack}(b) and the inset. 
At the peak, 
the GC breaks off and is divided into small clusters.
Note that the critical fraction $q_{c}$ is around $0.03$ 
for the SF network models \cite{Hayashi07}
and the Internet data at the level of autonomous system 
\cite{Albert00,Hayashi07,Satorras04}.
Figure \ref{fig_SN_failure}(a) shows 
that the values of $S(q)/N$ 
is sustained without a considerable drop against random node removals, 
except the case of $\delta=0.3$ (denoted by cyan dashed line).
Note that the virtual line of angle $-45^{\circ}$ 
is the most robust case of $R=0.5$ in the complete graph with the 
maximum $\langle k \rangle = N-1$: wasteful links. 
The critical fraction $q_{c}$ is about $0.8$ 
for the onion-like networks 
but $< 0.1$ for the tree-like networks 
as shown in Fig. \ref{fig_SN_failure}(b). 
Remember that 
the robustness index $R$ defined in Eq.(\ref{eq_def_index-R})
indicates the area under a line of $S(q)/N$.
Thus, 
higher robustness (values of $R$) in the growing onion-like networks 
is obtained than that in the tree-like networks for $\delta = 0.3$.
Note that the value of $R$ include more information in order to 
compare the robustness, 
since different values of $R$ can exits even when some 
networks have a same value of $q_{c}$.
The assortativity $r$ and robustness index $R$ 
are summarized in Table \ref{table_comparison_r-R}.
As the deletion rate $\delta$ is larger, 
$r$ is smaller, 
while $R$ is almost constant in both cases of 
the attacks and the failures.

\begin{table}
\newlength{\myheight}
\setlength{\myheight}{2.6mm}
\centering
\scalebox{1.3}{
\begin{tabular}{cc|ccc}
$\delta$ & $p_{sc}$ & $r$ & $R$ by attacks & $R$ by failures \\ \hline
\rule{-1mm}{\myheight}
0.4      & 0.003   & 0.412 & 0.246 & 0.427 \\
0.5      & 0.006   & 0.347 & 0.267 & 0.439 \\ 
0.6      & 0.010   & 0.277 & 0.283 & 0.447 \\
0.9      & 0.018   & 0.192 & 0.281 & 0.448 \\ 
0.3      & -       & 0.442 & 0.127 & 0.356 \\ \hline
\end{tabular}
}

\vspace{2mm}
\caption{Comparison of assortativity $r$ and robustness index $R$ 
in the networks 
for $N=5000$ with $\langle k \rangle \approx 6.3$.}
\label{table_comparison_r-R} 
\end{table}

\begin{figure}[htb]
\begin{center}
  \includegraphics[height=45mm]{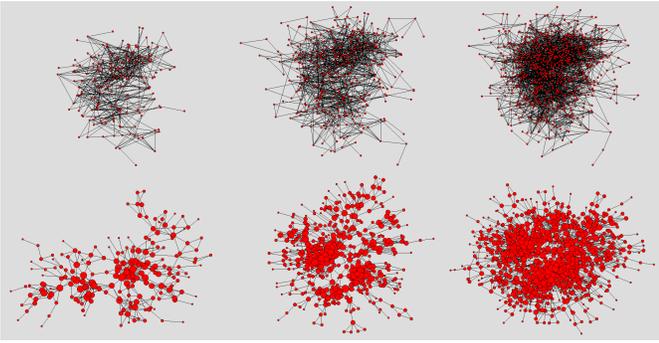}
\caption{Time evolution in the case of 
$\delta = 0.5$, $p_{sc}=0.006$, and $IT=50$
for $N = 200, 400$, and $800$ from left to right. 
(Top) Spatially growing networks. 
(Bottom) The corresponding onion-like topologies, in which the sizes of 
nodes are proportional to their degrees 
in order to show 
the core of connected high degree nodes 
and the periphery of connected low degree nodes.}
\label{fig_vis_evol}
\end{center}
\end{figure}

\begin{figure}[htb]
  \centering
   \includegraphics[height=95mm,angle=-90]{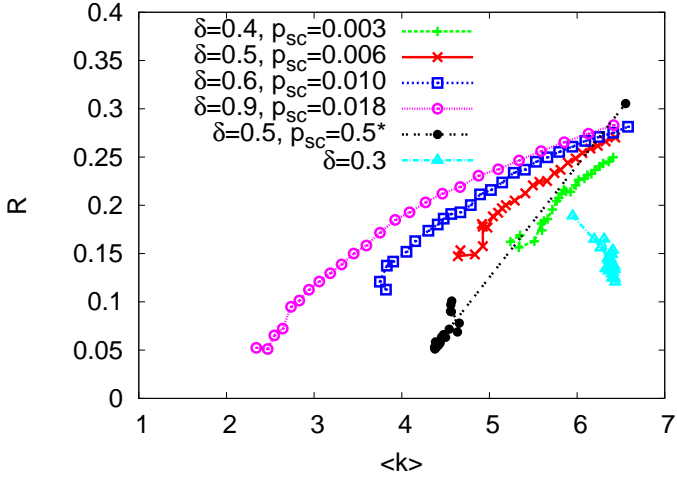}
\caption{Relation of the 
average degree $\langle k \rangle$ and the robustness index $R$
in the incrementally growing networks with $IT = 50$.}
\label{fig_map_R-k}
\end{figure}

We also study a spatially growing method. 
We consider 
the initial configuration of connected two nodes located at the 
center with the link length $20$ in a $160 \times 160$ square. 
At each time step in Step 1, 
from the center of a randomly chosen node, 
a new node is set on the radius of random number between $r_{min}$ 
and $r_{max}$ with any direction as similar to the basic idea of 
\cite{Brunet07}.
We set $r_{min}=15$ and $r_{max}=20$.
As shown in the top of Fig. \ref{fig_vis_evol}, 
the growing networks spread out diffusively on the space 
according to the time course.
Figure \ref{fig_map_R-k} shows the relation of 
$\langle k \rangle$ and $R$ through the growth. 
In the incrementally growing onion-like networks
for the values of $\delta$ and $p_{sc}$, 
the marks of green plus, red cross, blue open square, and 
magenta open circle on each line 
correspond to the size $N = 200, 400, 600, \ldots, 5000$ from 
left to right.
Therefore, both $\langle k \rangle$ and $R$ increase 
with the time course. 
The robustness is reinforced with increasing $\langle k \rangle$.
In other words, 
the network becomes more and more robust 
through the growth.
However, $\langle k \rangle$ hardly vary and 
$R$ slightly decreases in the tree-like networks 
denoted by the marks of cyan triangle and black filled circle.
The directions of lines are from top to bottom for 
 $N = 200, 400, 600, \ldots, 5000$ in the time course.
The long upper jump of a gray dashed line is due to become 
onion-like networks finally at $N=5000$ 
by adding many shortcut links with the rate $p_{sc}=0.5$ 
in Step 3.
Through the growth, 
the assortativity $r$ is almost constant around $0.3 \sim 0.4$, 
except it increases (from $-0.1$ to $0.2$) 
in the case of $\delta=0.9$ \& $p_{sc}=0.018$. 

\subsection{Efficiency for path lengths}
The efficiency for communication or transportation is measured by 
the average path lengths on the shortest paths over the network. 
The length is usually counted by the minimum number of hops 
between two nodes.
Because a path which passes 
through as few mediator nodes as possible gives 
lower cost (energy consumption) 
and more tolerant (safety or stable) operation for time-delay 
or information loss.
Short paths are 
also related to reduce traffic load defined by the betweenness 
centrality \cite{Freeman77}: 
the frequency of passing through a node or a link on the paths
between all pairs of nodes.
The frequency can be weighted by 
an inhomogeneously distributed (communication or transportation) 
requests, e.g. proportional to the products of populations around 
source and destination nodes on a space \cite{Hayashi12b, Dolev10}
in more realistic situation. 

\begin{figure}[htb]
\begin{center}
  \includegraphics[height=95mm,angle=-90]{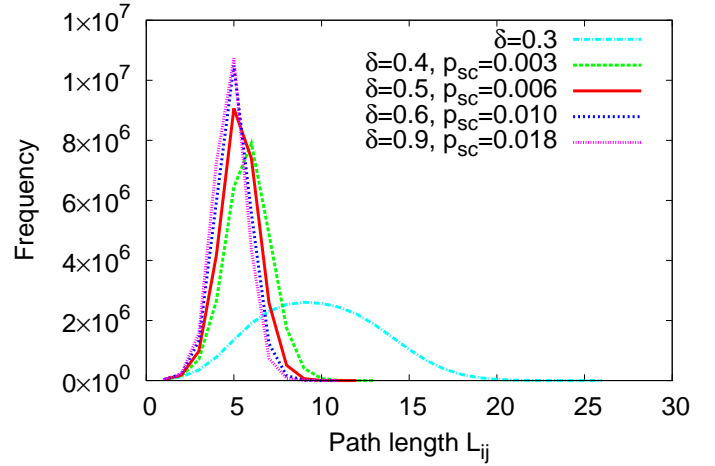}
\caption{Distribution of path lengths in the networks 
for $N=5000$ with $\langle k \rangle \approx 6.3$.
The case for $\delta = 0.3$ is tree-like networks, 
and the other cases are onion-like networks.}
\label{fig_dist_path-len}
\end{center}
\end{figure}

Figure \ref{fig_dist_path-len} shows that 
the distributions of path lengths 
in the networks at a same connection density level of 
$\langle k \rangle \approx 6.3$.
The widths are narrower around 5 hops 
in the growing onion-like networks denoted by green dashed, red solid, 
blue dashed, and magenta dotted lines for $\delta = 0.4$, $0.5$, $0.6$, 
and $0.9$ 
than that in the tree-like networks denoted by cyan chain line for 
$\delta = 0.3$.
The path lengths are short around 5 hops in average and 
at most about 10 hops for $N=5000$.
Thus, 
the growing onion-like networks are efficient with short paths.

\begin{figure}[htb]
\begin{center}
  \includegraphics[height=95mm,angle=-90]{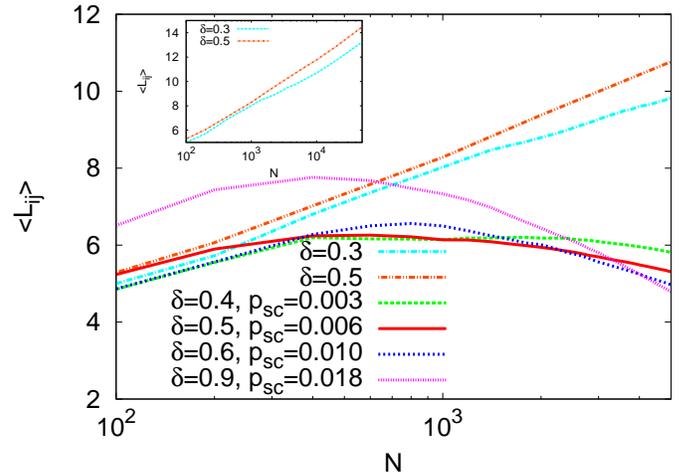}
\caption{Average path length $\langle L_{ij} \rangle$
counted by the number of minimum 
hops between two nodes over the network. 
The convex curves in the growing onion-like networks 
for $\delta = 0.4$, $0.5$, $0.6$, and $0.9$ 
are caused by the increase of $\langle k \rangle$
(see Fig \ref{fig_map_R-k}).
Inset show the SW property as a straight line in a semilog graph; 
the average path length on the tree-like networks for $\delta = 0.3$ 
or $0.5$ follows $O(\log N)$ in three-digit number.}
\label{fig_path-len_SW}
\end{center}
\end{figure}

In more detail for the emergence of SW property \cite{Watts98}, 
it is important whether the average path length follows 
$O(\log N)$ for the size $N$ under a constant connection density. 
Inset of Fig. \ref{fig_path-len_SW} shows the SW property in the 
tree-like networks 
denoted by cyan and orange chain lines for $\delta = 0.3$ and $0.5$.
Moreover, as shown in Fig. \ref{fig_path-len_SW}, 
the path lengths are very short around 5-6 hops 
in the growing onion-like networks denoted by green dashed, red solid, 
and blue dashed lines for $\delta = 0.4$, $0.5$, and $0.6$, 
except with slightly 
higher values of $\langle L_{ij} \rangle$ 
(longer path lengths) in $N < 10^{3}$ 
denoted by magenta dotted line for $\delta = 0.9$
than the case of tree-like networks 
denoted by cyan and orange chain lines for $\delta = 0.3$ and $0.5$.
We should remark that 
the average degree $\langle k \rangle$ 
is not constant within a connection density 
but increasing in the growing onion-like networks.
Moreover, $\langle k \rangle$ is smaller 
than that in 
the tree-like networks for $\delta = 0.3$ as shown in 
Fig. \ref{fig_map_R-k}. 
Smaller $\langle k \rangle$ means fewer total links, and 
the network construction is less expensive in lower cost 
especially in an early stage of growth.

Another important measure is the path distance, 
which is defined by the sum of Euclidean distances for the 
links on the shortest path (counted by the minimum number of hops)
between two nodes.
This measure is corresponding to 
the load of physical movements 
for (communication or transportation) flows on a space, 
instead of the load of routing process (for switching the 
transfer direction)
defined by the number of mediator nodes on the path.

\begin{figure}[htb]
\begin{center}
  \includegraphics[height=95mm,angle=-90]{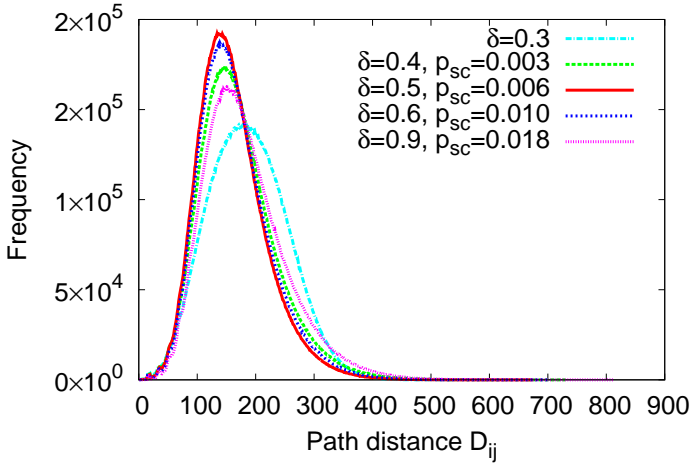}
\caption{Distribution of path distance in the networks 
for $N=5000$ with $\langle k \rangle \approx 6.3$.
The frequency is counted in the histogram of bin width one for 
real values of $D_{ij}$.
The case for $\delta = 0.3$ is tree-like networks, 
and the other cases are onion-like networks.}
\label{fig_dist_path-D}
\end{center}
\end{figure}

\begin{figure}[htb]
\begin{center}
  \includegraphics[height=95mm,angle=-90]{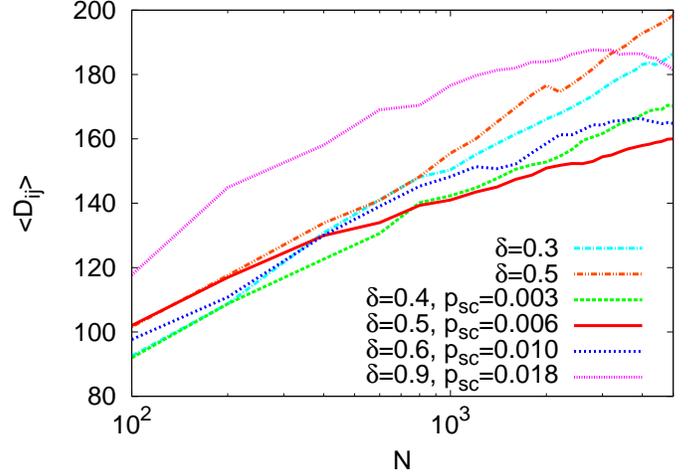}
\caption{Average path distance $\langle D_{ij} \rangle$
counted by the sum of Euclidean distances 
on the paths with the minimum hops 
between two nodes over the network.
The slopes of green, red, and blue lines in the onion-like networks 
are slightly gentle 
than ones of cyan and orange lines in the tree-like networks.}
\label{fig_path-D_SW}
\end{center}
\end{figure}

Figure \ref{fig_dist_path-D} shows the distribution of path distances 
$D_{ij}$, in which the majorities are short 
around the size $160$ of outer square. 
The case of $\delta = 0.3$ as the tree-like network 
denoted by cyan chain line has longer distances than other cases, 
since the distribution is shift to the right.
We also confirm the SW property of $O(\log N)$ for the path distance 
as shown in Fig. \ref{fig_path-D_SW}. 
The green dashed, red solid, and blue dotted lines 
for $\delta = 0.4$, $0.5$, and $0.6$ 
in the onion-like networks 
are below the cyan and orange chain lines for $\delta = 0.3$ and $0.5$ 
in the tree-like networks, 
except the magenta dotted line for $\delta = 0.9$.
Thus, the onion-like networks has shorter path distances 
than the tree-like networks within a same connection density level, 
although the advantage in the Euclidean distances 
is weak 
in comparison with that in the number of hops (see Figs. 
\ref{fig_path-len_SW} and \ref{fig_path-D_SW}).
Note that we have similar results for the path length $L_{ij}$ and 
distance $D_{ij}$, when we chose the shortest distance path 
instead of the above shortest path defined by 
the minimum number of hops between two nodes.
The difference of shortest (Euclidean) distance path and shortest 
(minimum hops) path 
is very small in these spatial networks, in other words, 
they give similar efficiency 
for the routing paths in these measures.

\section{Discussions for applications} \label{sec5}
In this section, we will discuss the possibilities or issues 
in our proposed networks for applications of 
communication or transportation systems. 
There exist flows which represent 
movements of one entity relative to another on a network. 
Thus, 
we categorize the dynamics in networks into the following types.

\begin{description}
  \item[Type 1. ] Dynamics of network configuration itself
  \item[Type 2. ] Dynamics of information flows, rumor spreading, 
    opinion formation, synchronization, or logistics on a network
\end{description}

Through this paper, 
we study the Type 1 of dynamics in order to aim 
a fundamental mechanism for 
the self-organized design of efficient and robust networks.
Temporal and/or fixed 
(corresponding to wireless and/or wired) connections are possible 
depending on the time-scale for changing the connection structure 
in a network. 
The quick change results in ad hoc networks, 
while the slow change is treated as an incremental 
modification of network. 
Both cases and the mixed one are not excluded, 
however we have assumed that each node or link is persisted 
once it is added unless removed by failures or attacks 
to simplify the discussion in Sections \ref{sec2}-\ref{sec4}.
The Type 2 of dynamics is significant for applications, e.g. 
wireless, sensor, mobile communication systems or 
autonomous transportation systems. 
In the Type 2 of dynamics, 
operation protocols for birth and death of communication or transportation 
request, routing, avoidance of congestion, 
task allocation, queuing, awareness of location, 
monitoring of system's states or conditions, and so on \cite{Barbeau07}, 
are necessary.
There are many methods for the efficient operations 
that are independent (generally applicable) or 
dependent on a special network structure. 
The detail discussions are strongly related to 
device technologies, users, and 
situations of utilization, 
they are beyond our current scope.

Even when we focus on the Type 1 of dynamics, there are 
two kinds of interactions among individuals (nodes) and with 
environment \cite{Dressler07} 
for the operation and control in a self-organized system. 
In particular, we consider 
changes of environment and the adaptivity in a network. 
Here, 
adaptivity is defined as 
{\it the capability to work in different or changing 
environment without intervention and configuration by an 
administrator} \cite{Dressler07}. 
The levels of adaptation are distinguished 
as shown in Table \ref{table_level_adaptation}.

\begin{table}
\centering
\scalebox{1.3}{
\begin{tabular}{ll} 
{\it Level 1}: & {\it The simplest adaptivity is to cope with changes}\\
         & {\it such as failures and mobility.}\\
{\it Level 2}: & {\it In order to react to changes and to optimize} \\
         & {\it the system performance, protocols should be} \\
         & {\it able to adapt their own parameters, e.g. timers} \\
         & {\it and cluster size.} \\
{\it Level 3}: & {\it In case of dramatic changes, the complete change} \\
         & {\it of the employed algorithms can be necessary, e.g.} \\
         & {\it the exchange of a no longer converging clustering} \\
         & {\it algorithm.} \\ 
\end{tabular}
}

\vspace{2mm}
\caption{Three level of adaptation quoted from pp.57 in Chapter 5 
of Ref.\cite{Dressler07}.}
\label{table_level_adaptation}
\end{table}

At the Level 1, 
a compensatory growth for removed parts by attacks or failures 
is related to an adaptivity with healing function 
in the incrementally growing onion-like networks 
because of the potential with strong robustness. 
The function of proxy at a new node locally 
contributes to make different 
access paths through the copying process 
as mentioned in subsection \ref{sec2} A.
On the other hand, 
node mobility may give rise to other problems 
of disconnections in a limited transmission range of wireless beam, 
handover for high-speed environment, or loss of data and navigational 
routes. 
When a communication network is often disconnected but resilient
due to node mobility, limited radio power, node or link failure, 
it is known as a {\it Delay/Disruption Tolerant Network} (DTN)
where a mobile device or software agent temporary stores
and carries local information
for forwarding messages until an end-to-end route is re-established 
or re-generated.
It is used in disasters, battlefields, and vehicular communications.
There are many protocols in the concept of 
DTN routing \cite{Shah11,DSouza10}.
As a base structure for DTN, 
it is obviously better to have many short paths and to maintain 
tolerant path connectivity for temporal changes of nodes or links. 
Our growing onion-like network  becomes one of the candidates of 
DTN. 

At the Level 2, in our growing network, 
it will be useful 
to regulate the parameters of $\delta$, $p_{sc}$, and $IT$ 
in order to repair the damaged parts and to recover 
the performance for communication or transportation 
according to a limited resource 
and the state of network system monitored at a time.
The scheduling is a further issue, and 
it is also important 
how to detect the damaged parts or malfunctions caused from 
overload beyond a capacity in a realistic network.
At the Level 3, a change of the intrinsic 
mechanism for generating a network is required. 
Since the required change 
depends on the scale or the speed of damages or evolutions, 
we should develop a proper measure to evaluate them in 
many complex situations. 

Apart from the basic property, 
only random failures and malicious attacks by removing nodes 
in decreasing order of their current degrees 
are insufficient to investigate 
the robustness in a realistic network.
Natural disasters such as earthquake or flood and unremitting 
armed conflicts with aerial bombing give rise to 
spatially spreading damages. 
They probably make complex shapes of removed part in the network 
according to a geographical map.
Although we treat 
a simple spatially growing method in Section \ref{sec4}, 
we need more consideration.
In addition, how to locate a node on a space should be discussed 
more carefully 
e.g. for the geographical allocation of 
mobile base stations for wireless communication in an 
emergent case of disaster or poor infrastructural environment.
In any cases, incremental repair and reinforced construction 
will be powerful approaches.

\section{Conclusion} \label{sec6}
We have proposed a self-organized design method for generating 
efficient and robust networks. It is based on biologically 
inspired copyings and the complementary adding shortcut links to 
enhance the robustness.
In particular, tanking into account positive degree-degree 
correlations, we have considered incrementally growing 
onion-like networks, which emerges 
from the backbone of tree-like structure generated by copyings 
and the periphery of shortcut links between low degree nodes. 
The obtained properties are fine as follows.
We emphasize that 
these properties are not trivially predicted from a combination 
of copyings and adding shortcuts.

\begin{itemize}
  \item non-singular self-averaging property unlike the conventional 
    D-D model \cite{Sole02, Satorras03, Ispolatov05} \\
    $\rightarrow$ to ensure statistically meaningful convergence 
    for a large size $N$ in samples
  \item growing with the compensatory function 
    of local proxy as another access point 
  \item exponential degree distribution without huge hubs\\
    $\rightarrow$ to avoid the concentration of linking cost and 
    traffic load at a few nodes \cite{Goh01}
  \item strong tolerance of connectivity against failures and attacks 
    equivalent to the nearly optimal in the rewired version \cite{Wu11} 
    under a same degree distribution 
  \item efficiency with short path lengths and distances 
    (counted by the number of hops and the Euclidean distances,
    respectively, on the shortest path) 
    for communication or transportation, 
    in addition the path length is 
    superior to the $O(\log N)$ SW property \cite{Watts98} in a sense 
\end{itemize}

Moreover, we have discussed the possibilities and issues for 
applications in wireless and/or wired communication or transportation 
systems, and particularly mentioned the adaptivity to heal over and 
to recover the performance of networking for a change of environment in 
such disasters or battlefields on a geographical map.
In this paper, 
we have considered a self-configuration of efficient and robust 
network in an incremental and distributed manner.
The proposed method does not require 
the entirely rewiring processes on a network \cite{Wu11}. 
Thus, the obtained results 
will be useful for temporal evolution of a resilient network 
system.
Other capabilities and realistic strategies for 
self-diagnosis, self-protection, 
self-healing, self-repair, and self-optimization \cite{Dressler07} 
are further issues.
We will study effectively applicable 
methods to self-organize networks which 
keep high robustness and efficient paths for communication 
or transportation in both normal and emergent situations.


\section*{Acknowledgment}
This research is supported in part by
a Grant-in-Aid for Scientific Research in Japan, No. 25330100.



%

\end{document}